\documentclass[%
 reprint,
superscriptaddress,
nofootinbib,
 amsmath,amssymb,
 aps,
 prd,
]{revtex4-1}

\usepackage{lineno,hyperref}

\usepackage{graphicx}
\usepackage{dcolumn}
\usepackage{bm}

\begin{document}

\date{\today}
\title{Detecting the Diffuse Supernova Neutrino Background with LENA}

\author{R. M\"{o}llenberg}
\email{randolph.moellenberg@ph.tum.de}
\affiliation{Excellence Cluster Universe, Technische Universit\"{a}t M\"{u}nchen, 85748 Garching, Germany}
\author{F. von Feilitzsch}
\affiliation{Physik Department, Technische Universit\"{a}t M\"{u}nchen, 85748 Garching, Germany}
\author{D. Hellgartner}
\affiliation{Physik Department, Technische Universit\"{a}t M\"{u}nchen, 85748 Garching, Germany}
\author{L. Oberauer}
\affiliation{Physik Department, Technische Universit\"{a}t M\"{u}nchen, 85748 Garching, Germany}
\author{M. Tippmann}
\affiliation{Physik Department, Technische Universit\"{a}t M\"{u}nchen, 85748 Garching, Germany}
\author{J. Winter}
\affiliation{Institut f\"{u}r Physik, Excellence Cluster PRISMA, Johannes Gutenberg Universit\"{a}t Mainz, 55128 Mainz, Germany}
\author{M. Wurm}
\affiliation{Institut f\"{u}r Physik, Excellence Cluster PRISMA, Johannes Gutenberg Universit\"{a}t Mainz, 55128 Mainz, Germany}
\author{V. Zimmer}
\affiliation{Physik Department, Technische Universit\"{a}t M\"{u}nchen, 85748 Garching, Germany}

\begin{abstract}

LENA (\textbf{L}ow \textbf{E}nergy \textbf{N}eutrino \textbf{A}stronomy) has been proposed as a next generation
50\,kt liquid scintillator detector. Its large target mass allows to search for the Diffuse Supernova Neutrino Background (DSNB),
which was generated by the cumulative emissions of all core-collapse supernovae throughout the universe. Indistinguishable background from reactor and
atmospheric electron antineutrinos limits the detection window to the energy range between 9.5\,MeV and 25\,MeV. Depending on the mean supernova neutrino energy, about
5 to 10 events per year are expected in this energy window. The background from neutral current reactions of atmospheric neutrinos surpasses
the DSNB by more than one order magnitude, but can be suppressed by pulse shape discrimination. Assuming that the residual background is
known with 5\,\% uncertainty, the DSNB can be detected with $3\,\sigma$ significance after 10 years of data taking. In case that no hint for a signal is seen,
current standard DSNB models would be ruled out with more than 90\,\% C.L. 

\end{abstract}

\maketitle


\section{Introduction}

The cumulative neutrino emissions of all core-collapse supernovae throughout the universe have generated the Diffuse Supernova Neutrino Background (DSNB).
The DNSB contains information about the redshift dependent core-collapse supernova rate and about the average neutrino spectrum of a core-collapse
supernova. Up to now the DSNB could not be detected due to the low flux. The best limit was achieved by the Super-Kamiokande experiment
which sets an upper limit (90\,\% C.L.) on the flux of $\mathrm{3.1\,\bar\nu_e\, cm^{-2}s^{-1}}$ for $\mathrm{E_{\bar\nu_e}>17.3\,MeV}$ \cite{superK_dsnb}, which is about a factor of two to four above the currently predicted values \cite{superK_dsnb}.

The DSNB contains neutrinos and antineutrinos of all flavours, but the detection channel with the largest cross section, the inverse beta
decay channel ($\mathrm{\bar\nu_e + p \rightarrow e^{+} +n }$), is only sensitive to electron antineutrinos. 
Compared to Super-Kamiokande, the proposed LENA (\textbf{L}ow \textbf{E}nergy \textbf{N}eutrino \textbf{A}stronomy) detector \cite{lenawhitepaper} has the advantage that it can detect the 2.2\,MeV gamma of the neutron capture
on hydrogen. Hence, the coincidence of the prompt positron and the delayed neutron signal can be used to suppress background events, which is crucial to detect
the low DSNB flux.

The detection of the DSNB in LENA was analyzed in \cite{dsnbpaper}, neglecting the background from neutral current reactions of atmospheric neutrinos,
which was observed by the KamLAND experiment \cite{kamland_nc_bg} after the publication of \cite{dsnbpaper}.
In the present work, this important background is included for the first time and the calculations of the other backgrounds
are updated. In Sec.\ \ref{sec:det} a brief description of the LENA detector is presented.
The simulation of the DSNB and the corresponding background spectra is discussed in Sec.\ \ref{sec:dsnb_sim} and Sec.\ \ref{sec:bg}. As the background
from neutral current reactions of atmospheric neutrinos surpasses the DSNB signal \cite{kamland_nc_bg}, an efficient reduction of this background is neccessary.
Sec.\ \ref{sec:nc_psd} discusses how this can be achieved by pulse shape discrimination. Finally, the detection potential of the DSNB is discussed in Sec.\
\ref{sec:dsnb_detection}.

\section{The LENA Detector}
\label{sec:det}

\begin{figure}[!htbp]\centering\includegraphics[width=0.45\textwidth]{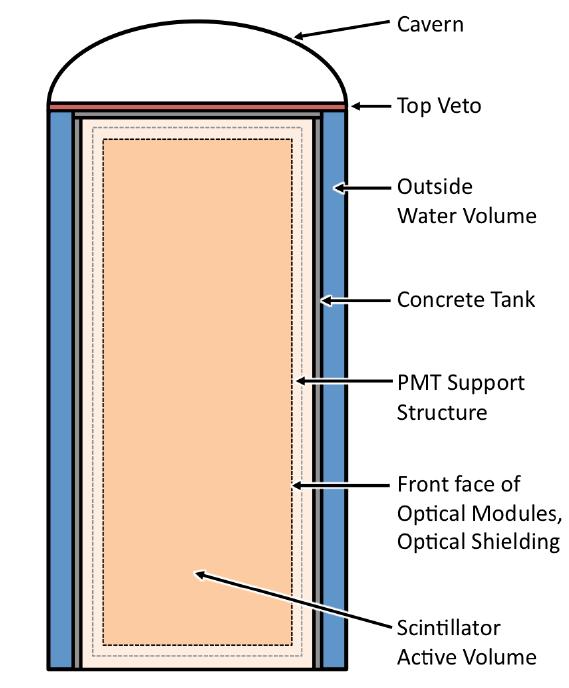}
\caption[Schematical view of the \textsc{LENA} detector]{Schematical view of the \textsc{LENA} detector \cite{lenaspec}.   
}
\label{fig:lena}
\end{figure}

Figure \ref{fig:lena} shows a schematic overview of the currently proposed LENA design \cite{lenaspec}. The neutrino target consists of $\sim$50\,kt 
of liquid scintillator based on linear-akyl-benzene (LAB). The emitted light is detected by photomultiplier tubes (PMTs)
that are mounted with non-imaging light concentrators (LCs) inside individual pressure encapsulations
that are filled with a non-scintillating buffer liquid. The apertures of these optical modules face the boundary of the
target volume at a radius of 14\,m. The corresponding effective optical coverage is $\sim30\,\%$. The radius of the cylindrical concrete tank is 16\,m, so that the target volume is shielded by 2\,m of liquid scintillator. A muon veto formed by gas detectors is placed above the detector tank and provides auxiliary information for the reconstruction of cosmic muon tracks. In order to identify and reconstruct also inclined muon tracks, an instrumented water volume surrounding the tank serves as an active 
Water-$ \mathrm{\check C}$erenkov muon veto. Additionally, it shields the target volume from fast neutrons.

The preferred location for the detector is the Pyh\"{a}salmi mine in Finland, that has 1400\,m of rock coverage, corresponding to
a shielding of 4000\,m water equivalent (w.e.).
Hence, the cosmic muon flux is reduced to $\sim\mathrm{0.2\,m^{-2}h^{-1}}$ \cite{fastneutronsim}.

\section{Simulation of the Diffuse Supernova Neutrino Background Spectrum}
\label{sec:dsnb_sim}

The detection channel for the DSNB is the inverse beta decay (IBD) reaction of an $\bar\nu_e$ on a free proton
($\mathrm{\bar\nu_e + p \rightarrow e^{+} +n }$), as it has the largest cross section at low energies.
Due to the kinematics of the IBD reaction, the kinetic energy of the neutron can be neglected \cite{IBD_Xsec}. Thus, the kinetic
energy of the positron is almost equal to the incident neutrino energy, reduced by the 1.8\,MeV Q-Value of the IBD reaction.
Hence, the neutrino energy can be reconstructed from the detected visible energy ($\mathrm{E_{\nu} \approx E_{vis}+0.8\,MeV}$), considering
that the positron annihilates with an electron into two 511\,keV gammas.
The produced neutron gets captured by a proton with a mean capture time of $\sim250\mu s$, producing a 2.2\,MeV gamma. Hence, the delayed
coincidence between the prompt positron and the delayed neutron signal can be used to reject background events.

Assuming that the supernova neutrino spectrum and the luminosity is independent of the neutrino flavour
such that $\mathrm{\langle E_{\bar\nu_e} \rangle=\langle E_{\bar\nu_\mu} \rangle=\langle E_{\bar\nu_\tau} \rangle}$
and $\mathrm{L_{\bar\nu_e}=L_{\bar\nu_\mu}=L_{\bar\nu_\tau}=0.5\cdot10^{53}erg}$,
neutrino oscillations can be neglected.

Thus, the detected DSNB spectrum is given by:
\begin{equation}
\mathrm{\frac{dR_{\nu}}{dE_{\nu}}=\frac{dF_{\nu}}{dE_{\nu}}\cdot \sigma_{\nu}(E_{\nu}) \cdot N_p \ ,}
\label{eq:dsnb_spectrum}
\end{equation}

where $\mathrm{\frac{dF_{\nu}}{dE_{\nu}}}$ is the DSNB flux, $\mathrm{\sigma_{\nu}(E_{\nu})}$
is the energy dependent cross section of the IBD reaction \cite{IBD_Xsec} and $\mathrm{N_p=3.67\cdot10^{33}}$ is the number of protons in the target volume
($\mathrm{r<13.5\,m}$, $\mathrm{|z|<48.5\,m}$)\footnote{Note that the volume directly infront of the optical modules can not be used
due to a very inhomogeneous detector response in this area \cite{moellenbergphd}.} \cite{teresaphd}.

The DSNB flux is calculated with the following expression \cite{dsnb_theory}:
\begin{equation}
\mathrm{\frac{dF_{\nu}}{dE_{\nu}}=\frac{c}{H_0}\int_0^{z_{m}}  \frac{dN_{\nu}(E^{'}_{\nu})}{dE^{'}_{\nu}}}
\mathrm{\frac{R_{SN}(z) dz}{\sqrt{\Omega_{m}\left(1+z\right)^{3}+\Omega_{\Lambda}}} }
\label{eq:dsnbflux}
\end{equation}
where $\mathrm{R_{SN}(z)}$ is the supernova rate at redshift z, $\mathrm{E^{'}_\nu=E_\nu(1+z)}$ is the redshift corrected energy of the neutrinos,
$\mathrm{\frac{dN_{\nu}(E_{\nu})}{dE_{\nu}}}$ is the mean number spectrum of the neutrinos emitted by one supernova explosion, $\mathrm{H_0}$
is the Hubble constant, c is the speed of light, $\mathrm{\Omega_m}$ is the cosmic matter density, $\Omega_{\Lambda}$ is the cosmic
constant and $\mathrm{z_{m}}=5$ is the redshift where the first core-collapse supernova occurred.

Assuming that every star above 8 solar masses ends in a core-collapse supernova and using
the Salpeter initial mass function \cite{salpeter}, $\mathrm{R_{SN}(z)}$ can be derived from the star formation rate $\mathrm{R_{\ast}(z)}$ to be
$\mathrm{R_{SN}(z)= 0.0122\,M^{-1}_{\odot} R_{\ast}(z)}$.

The used parametrization for the star formation rate is \cite{dsnb_theory}:
\begin{equation}
\begin{split}
\mathrm{R_{\ast}(z)=0.32 f_{SN} \frac{H_0}{70\,km s^{-1} Mpc^{-1}} \frac{e^{3.4z}}{e^{3.8z}+45}} \\
\mathrm{\cdot\frac{\sqrt{\Omega_{m}\left(1+z\right)^{3}+\Omega_{\Lambda}}}{(1+z)^{3/2}} yr^{-1}Mpc^{-3}} \ ,
\end{split}
\label{eq:sfr}
\end{equation}
where $\mathrm{f_{SN}=1.5\pm0.3}$ is a normalization factor, such that $\mathrm{R_{SN}(z=0)=1.25\cdot10^{-4}y^{-1}Mpc^{-3}}$ \cite{dsnb_beacom}.

The neutrino emission spectrum $\mathrm{\frac{dN_{\nu}(E_{\nu})}{dE_{\nu}}}$ is approximated by a Maxwell-Boltzmann spectrum \cite{lena_whitepaper}. 
The assumed mean supernova neutrino energy $\mathrm{\langle E_\nu \rangle}$ is in the range between 12\,MeV and 21\,MeV, according
to indirect constraints from chemical abundances of neutrino induced elements \cite{sn_nucleo_synth},
which are in agreement with numerical supernova simulations \cite{krjsnmodel,tbpsnmodel, sn_model_sumiyoshi}.

Using equation \ref{eq:dsnb_spectrum}, four different DSNB spectra with $\mathrm{\langle E_\nu \rangle}$ ranging from 12\,MeV to 21\,MeV
were obtained from a GEANT4 \cite{geant4simtoolkit} based Monte Carlo simulation of the LENA detector \cite{moellenbergphd}. The events were homogeneously
distributed over the target volume, so that possible position dependent effects are considered.

\begin{figure}[!htbp]\centering\includegraphics[width=0.49\textwidth]{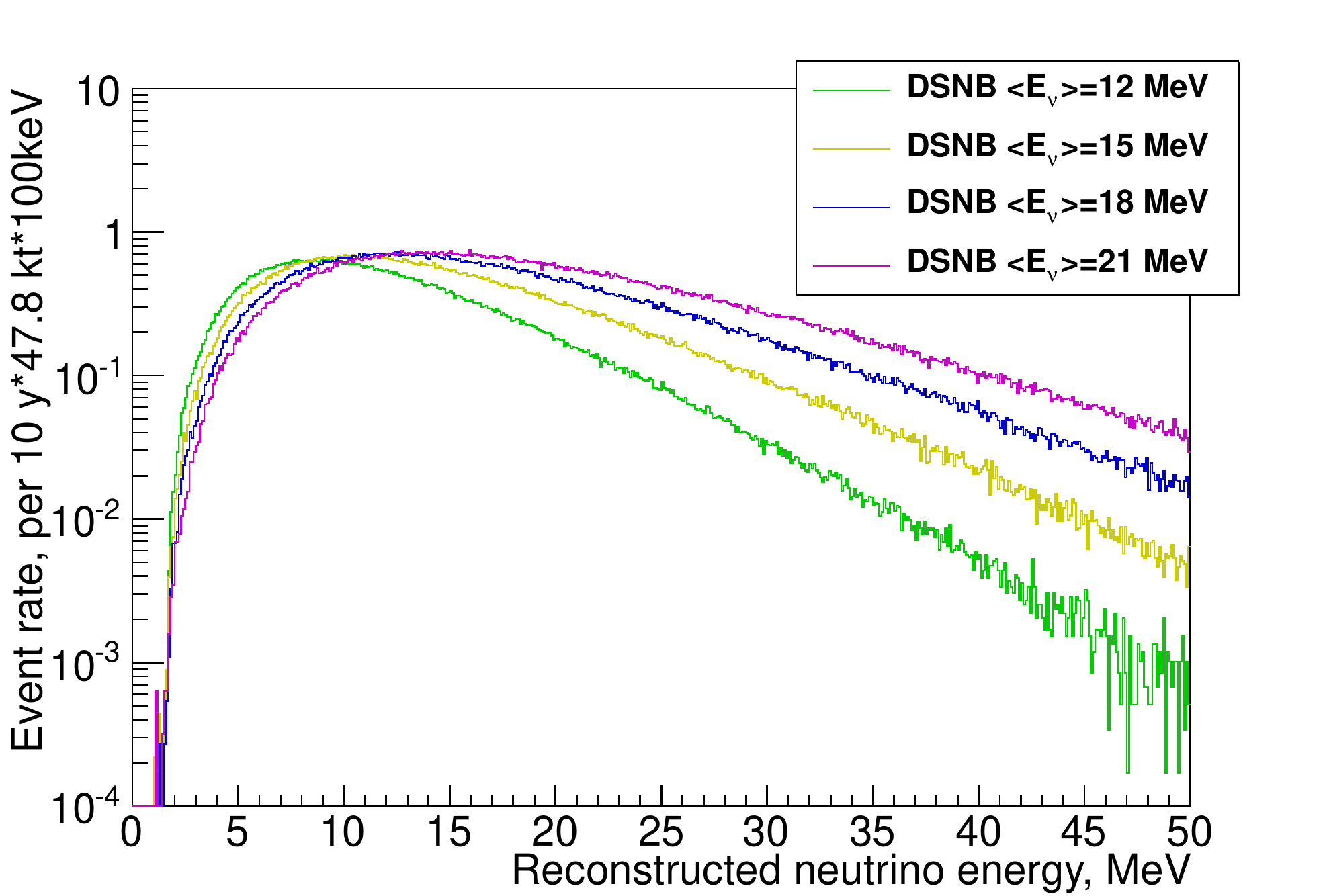} 
\caption[DSNB spectra]{
The expected DSNB spectra in LENA for different supernova neutrino spectra with mean energies ranging from 12\,MeV to 21\,MeV.}
\label{fig:dsnb_spectra}
\end{figure} 

Figure \ref{fig:dsnb_spectra} shows the resulting DSNB spectra. The spectra peak at energies between $\sim8\,$MeV and $\sim16\,$MeV, depending
on $\mathrm{\langle E_\nu \rangle}$ and decrease exponentially at higher energies.
The energy of the peak is lower than the corresponding $\mathrm{\langle E_\nu \rangle}$,
because the neutrino energy is redshifted. After 10 years of data taking, about 80 to 150 DSNB events are expected. Thus, very low
background levels are needed for a detection of the DSNB.

\section{Backgrounds}
\label{sec:bg}

The background for the DSNB detection can be divided into three categories:
Indistinguishable background from other $\mathrm{\bar\nu_e}$ sources, muon induced backgrounds and the neutral current reactions of
atmospheric neutrinos.

\subsection{$\mathrm{\bar\nu_e}$ Charged Current Background}

Other $\mathrm{\bar\nu_e}$ sources in the relevant energy region pose a dangereous background for the DSNB detection, as these events are indistinguishable
from DSNB events. Possible $\mathrm{\bar\nu_e}$ sources are reactor neutrinos and atmospheric neutrinos.

\begin{figure}[!htbp]\centering\includegraphics[width=0.49\textwidth]{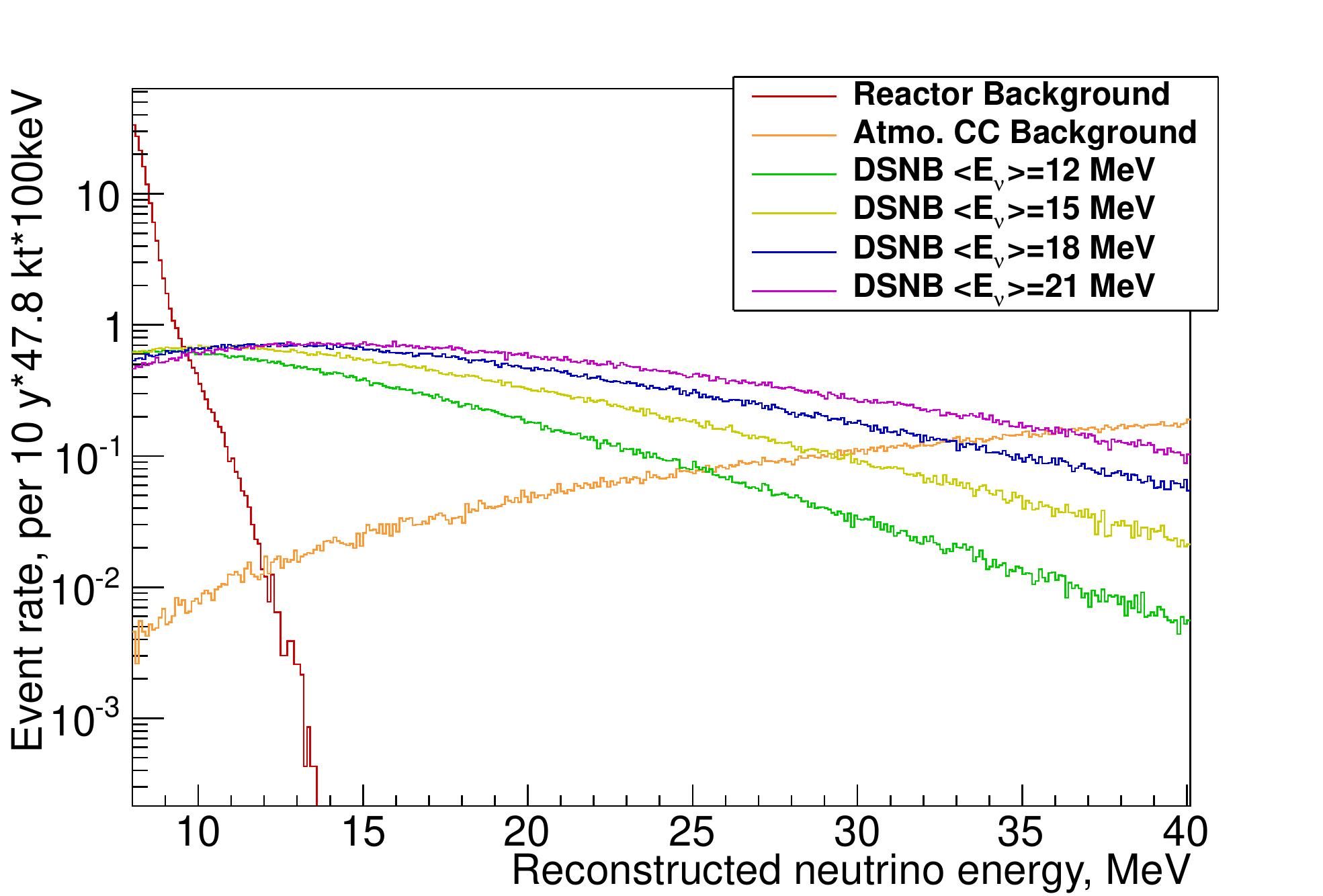} 
\caption[Reactor and atmospheric $\mathrm{\bar\nu_e}$ background]{
The reactor and atmospheric $\mathrm{\bar\nu_e}$ spectrum expected for LENA at Pyh\"{a}salmi. 
For comparison, the DSNB spectra with $\mathrm{\langle E_\nu \rangle}$ ranging from 12\,MeV to 21\,MeV are
depicted. }
\label{fig:atmo_cc_spectrum}
\end{figure} 

The reactor and atmospheric $\mathrm{\bar\nu_e}$ flux at the Pyh\"{a}salmi location were calculated in \cite{wurmphd}. Based on these results, the reactor and
atmospheric $\mathrm{\bar\nu_e}$ spectra were simulated analogous to the DSNB signal.
Figure \ref{fig:atmo_cc_spectrum} shows the resulting reactor and atmospheric background spectra. Below about 9.5\,MeV, the reactor $\mathrm{\bar\nu_e}$
spectrum surpasses the DSNB spectra and decreases exponentially with rising energy. The atmospheric $\mathrm{\bar\nu_e}$ background is suppressed at low energies
and rises with the energy until it surpasses the DSNB spectrum for $\mathrm{\langle E_\nu \rangle}=12$\,MeV at $\sim25$\,MeV reconstructed neutrino energy.
Hence, the detection window is set to 9.5\,MeV-25\,MeV, which further reduces the event rate to between 50 and 100 events
in 10\,y. About 11 remaining reactor and atmospheric $\bar\nu_e$ events are expected in the detection window.

\subsection{Muon-induced Backgrounds}
\label{sec:muon-induced}

When a cosmic muon traverses the target volume,
it can produce radioactive isotopes by spallation reactions on carbon. The majority of these radioisotopes can be vetoed by the delayed coincidence
condition. However, $^9$Li ($\mathrm{Q_{\beta^-}=13.6\,MeV}$) can $\beta^-$ decay into excited states of $^{9}$Be, leading to the emission of a neutron \cite{tableofisotopes}.
About $10^3$ $^9$Li $\mathrm{\beta-n}$ events are expected per 10\,y in the detection window. $^9$Li has a short life time
(257.2\,ms) and is produced close to the muon track. Thus, the $^9$Li background can be reduced to less than 0.01 events per 10\,y, by vetoing a cylinder
with 2\,m radius around each muon track for 2.5\,s which introduces 0.2\,\% effective dead time \cite{moellenbergphd}.

Another possible background are fast neutrons which are produced in the surrounding rock by cosmic muons. A small fraction of these
neutrons propagate into the target volume, without triggering the Water-$ \mathrm{\check C}$erenkov muon veto.
Inside the target volume, the neutrons cause a prompt signal by scattering reactions
on hydrogen and carbon. Afterwards, they thermalize and are finally captured by a proton, thus mimicking the delayed coincidence signature.
The fast neutron background was simulated with the GEANT4-based Monte Carlo simulation, assuming a mean cosmic muon energy of 300\,GeV \cite{fastneutronsim}.
It was found that the fast neutron rate in the target volume is about one order of magnitude larger than the DSNB event rate.
But as the fast neutrons enter the detector from outside, they are not homogeneously distributed over the target volume.
Instead, the majority of the fast neutrons are events close to the 
boundary of the target volume. Thus, it is possible to lower the fast neutron background to $\mathrm{4.9 \pm 0.4 (stat.)}$ events per 10\,y, by reducing
the fiducial volume radius to 11.0\,m. This cut reduces the DSNB event rate by 34\,\%, so that only 32.4 events per 10\,y are expected for
$\mathrm{\langle E_\nu \rangle}=12$\,MeV. Another possible method to reduce this background is by pulse shape analysis (see Sec. \ref{sec:nc_psd}).

\subsection{Neutral Current Atmospheric $\nu$ Background}

Additional to the CC background of atmospheric $\mathrm{\bar\nu_e}$, NC reactions of atmospheric neutrinos and antineutrinos of all
flavours pose a background for the DSNB detection. This NC background was first observed by the KamLAND experiment \cite{kamland_nc_bg} and was not included in previous analyses \cite{dsnbpaper,wurmphd}.
Several neutrino reactions contribute to this background. They all have in common that a single neutron is emitted
(e.g. $\mathrm{\nu_x + {^{12}C} \rightarrow \nu_x + n + {^{11}C} }$, though other more complex reactions are also possible), which mimics the 
IBD event signature (see Section \ref{sec:muon-induced}). 

\begin{figure}[!htbp]\centering\includegraphics[width=0.49\textwidth]{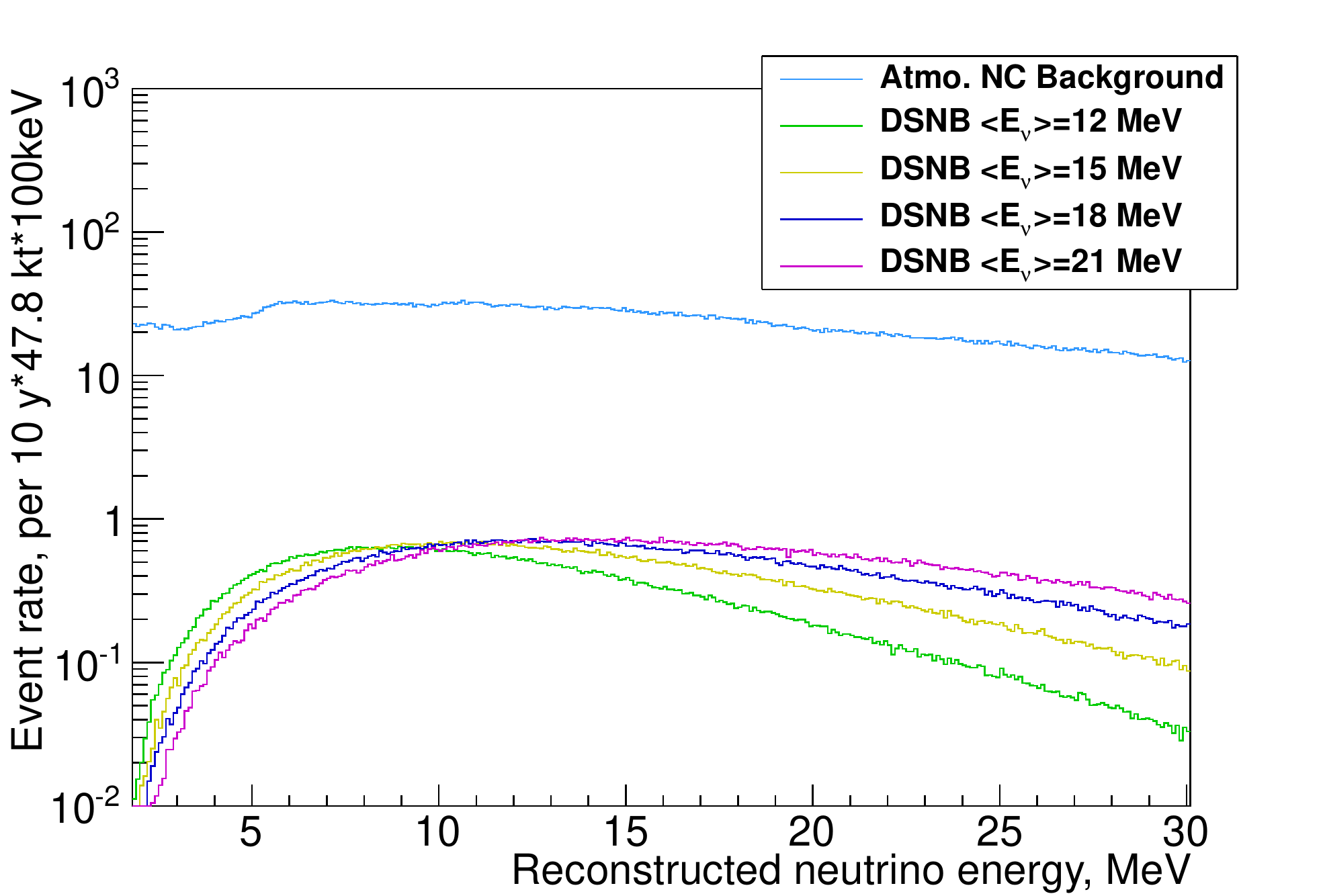} 
\caption[Atmospheric neutrino NC background spectrum]{
The simulated atmospheric neutrino NC background spectrum.
For comparison, the DSNB spectra with $\mathrm{\langle E_\nu \rangle}$ ranging from 12\,MeV to 21\,MeV are also
depicted.
}
\label{fig:atmo_nc_bg}
\end{figure}

In order to simulate this background, the reactions of atmospheric neutrinos inside the target volume were simulated
with the GENIE Neutrino Monte-Carlo Generator (Version 2.6.6) \cite{Genie}, using the Bartol atmospheric neutrino fluxes \cite{atmo_flux} as input. As these fluxes were calculated for the Super-Kamiokande experiment,
the resulting event rates have to be scaled by a factor of two in order to consider the higher geographic
latitude of the Pyh\"{a}salmi location \cite{wurmphd}.
After simulating the neutrino interaction, the final state particles were tracked with the GEANT4-based simulation of the LENA detector. Figure \ref{fig:atmo_nc_bg} shows the resulting atmospheric NC background spectrum.

The NC background spectrum surpasses the expected DSNB spectra by more than one order of magnitude over the whole energy region. 
Overall, $3.27\cdot 10^3$ events per 10\,y are expected in the DSNB detection window\footnote{Considering the larger atmospheric neutrino flux, this
result is in agreement with the KamLAND measurement \cite{kamland_nc_bg}.}. 
Hence, an efficient method to supress this background is necessary in order to detect the DSNB.

\begin{table}
\begin{center}
\begin{tabular}{|l|c|}
\hline
Reaction channel       &  Branching ratio\\                       
\hline
(1) $\mathrm{\nu_x+ {^{12}C}\rightarrow \nu_x + n + {^{11}C}} $ &38.8\,\% \\
(2) $\mathrm{\nu_x+ {^{12}C}\rightarrow \nu_x + p+n + {^{10}B}} $ &20.4\,\% \\
(3) $\mathrm{\nu_x+ {^{12}C}\rightarrow \nu_x + 2\,p+n + {^{9}Be}} $ &15.9\,\% \\
(4) $\mathrm{\nu_x+ {^{12}C}\rightarrow \nu_x + p+d+n + {^{8}Be}} $ &7.1\,\% \\
(5) $\mathrm{\nu_x+ {^{12}C}\rightarrow \nu_x + \alpha+p +n+ {^{6}Li}} $ &6.6\,\% \\
(6) $\mathrm{\nu_x+ {^{12}C}\rightarrow \nu_x + 2\,p+d+n+{^{7}Li}} $ &1.3\,\% \\
(7) $\mathrm{\nu_x+ {^{12}C}\rightarrow \nu_x + 3\,p+2\,n+{^{7}Li}} $ &1.2\,\% \\
(8) $\mathrm{\nu_x+ {^{12}C}\rightarrow \nu_x + d+n+{^{9}B}} $ &1.2\,\% \\
(9) $\mathrm{\nu_x+ {^{12}C}\rightarrow \nu_x + 2\,p+t +n+ {^{6}Li}} $ &1.1\,\% \\
(10) $\mathrm{\nu_x+ {^{12}C}\rightarrow \nu_x + \alpha+n+{^{7}Be}} $ &1.1\,\% \\
(11) $\mathrm{\nu_x+ {^{12}C}\rightarrow \nu_x + 3\,p+n+{^{8}Li}} $ &1.1\,\% \\
\hline
other reaction channels & 4.2\,\%\\
\hline
\end{tabular}
\end{center}
\caption[The branching ratios of atmospheric neutrino NC background reaction channels]
{The branching ratios of the different atmospheric NC background channels in the DSNB detection window.}
\label{tab:br_ratio_nc_bg}
\end{table}

A possible background suppression method is to look for the coincidence of an atmospheric neutrino NC event with the subsequent decay of 
any produced radioactive isotopes. Table \ref{tab:br_ratio_nc_bg} shows the branching ratios of the different atmospheric NC
background channels in the DSNB detection window. $^{11}$C is produced with a branching ratio of about 39\,\%. In this case it is
possible to veto the atmospheric NC event by looking for the coincidence of the IBD-like event and the subsequent decay of $^{11}$C.
As the lifetime of $^{11}$C is $\mathrm{\tau=29.4\,min}$, this cut can only be applied if the fiducial volume is reduced to 30\,kt,
in order to prevent accidental coincidences of IBD events and background events from external gamma rays.
But except $^8$Li which is produced in reaction (11), the other produced isotopes are either stable ($\mathrm{^{10}B}$, $\mathrm{^{9}Be}$, $\mathrm{^{7}Li}$, $\mathrm{^{6}Li}$), have a too long life time ($\mathrm{^{7}Be}$, $\mathrm{\tau=76.9\,d}$) or decay almost instantly ($\mathrm{^{9}B}$, $\mathrm{^{8}Be}$).

Hence, the atmospheric NC background can only be reduced by about 40\,\% by looking for the coincident decay of a radioactive isotope. 
Thus, a more efficient background suppression method is needed. 
A possible option is to reject atmospheric neutrino NC events by analyzing their pulse shape. This method will be discussed in the next Section.

\section{Pulse Shape Discrimination of Background Events}
\label{sec:nc_psd}

\begin{figure}[!htbp]\centering\includegraphics[width=0.49\textwidth]{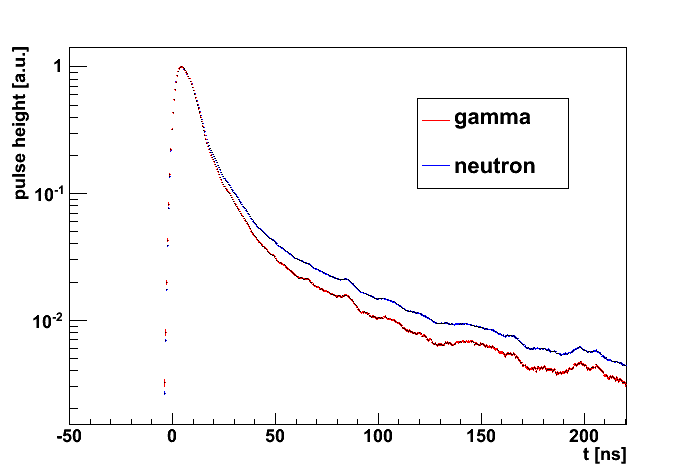} 
\caption[Average gamma and neutron pulse shape]{
Average normalized pulses for gamma (denoted in red) and neutron (denoted in blue) events \cite{juergenphd}.
}
\label{fig:average_ps}
\end{figure}

Heavier particles, like protons, neutrons or alphas, have a different pulse shape than positrons.
Figure \ref{fig:average_ps} shows a comparison between the average neutron and gamma pulse shape as measured in a small
scale laboratory experiment at the Maier-Leibnitz-Laboratorium (MLL) in Garching, Germany \cite{juergenphd}.
A clear difference between the two pulse shapes is visible. Hence, this difference can be used to distinguish neutron from positron
events. The neutron pulse shape was parametrized by the following PDF:
\begin{equation}
\mathrm{F(t)=\sum_{i}\frac{N_i}{\tau_i} e^{-\frac{t}{\tau_i}}} \ ,
\end{equation}
which was used as an input value for Monte-Carlo simulation of the LENA detector.
Table \ref{tab:proton_pulse_shape} shows the parameters of the photon emission PDF that were used for the simulation
of electrons, positrons, protons and alphas. The used scattering and absorption lengths are denoted in table \ref{tab:photonpropagation}.

\begin{table}
\begin{center}
\begin{tabular}{|c|c|c|c|}
\hline
Parameter & Electrons/Positrons & Protons & Alphas\\
\hline
$\mathrm{N_1}$ & 0.67 & 0.61& 0.44\\
$\mathrm{N_2}$ & 0.19 & 0.21& 0.16\\
$\mathrm{N_3}$ & 0.14 & 0.18& 0.40\\
$\tau_1$  & 6.8\,ns & 7.0\,ns & 3.2\,ns\\
$\tau_2$  & 26.5\,ns & 27.3\,ns & 18\,ns\\
$\tau_3$  & 152.3\,ns & 140.3\,ns & 190\,ns\\
$\mathrm{k_b}$ & $\mathrm{0.15\frac{mm}{MeV}}$ & $\mathrm{0.12\frac{mm}{MeV}}$ & $\mathrm{0.11\frac{mm}{MeV}}$\\
\hline
\end{tabular}
\end{center}
\caption[Photo emission parameters for electrons and protons]
{The photon emission parameters for electrons, protons and alphas \cite{juergenphd,labpulseshape,kblab}. The $\mathrm{k_b}$ value for 
protons was taken from a calibration measurement in the Borexino experiment, which uses PC as scintillator \cite{kb_proton}.}
\label{tab:proton_pulse_shape}
\end{table}

\begin{table}
\begin{center}
\begin{tabular}{|c|c|}
\hline
Parameter & Value \\
\hline
Rayleigh scattering length & 40\,m\\
Absorption-reemission length & 60\,m\\
Absorption length & 20\,m\\
\hline
\end{tabular}
\end{center}
\caption[Scattering and absorption lengths]{The values for the scattering and absorption lengths \cite{wurmphd} which were used in the simulation.
As the absorption length can not be measured directly, it was set to a value
that is in agreement with the attenuation length measurements that were performed in \cite{attenuation_length}.}
\label{tab:photonpropagation}
\end{table}

\begin{figure}[!htbp]\centering\includegraphics[width=0.5\textwidth]{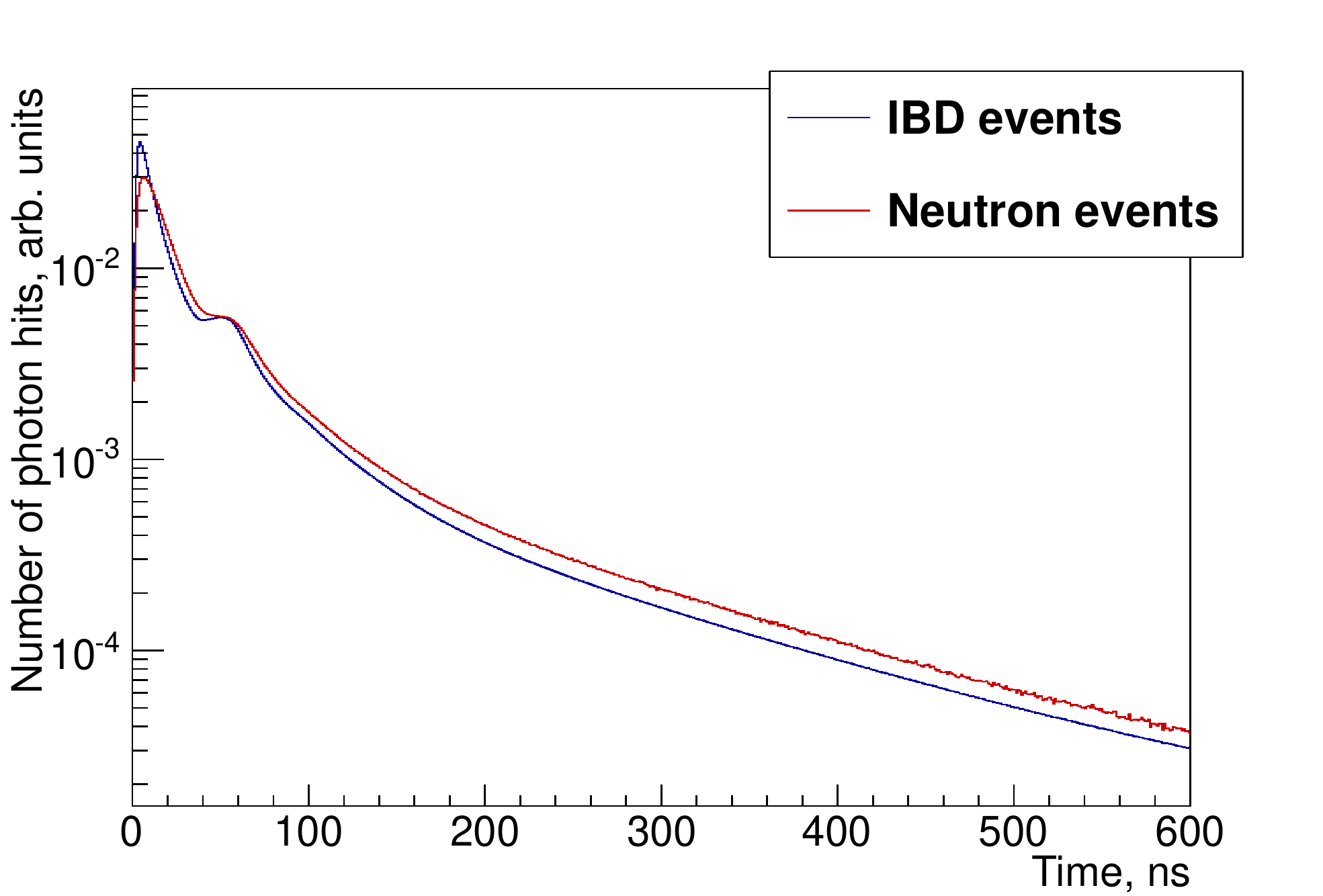}
\caption[The average positron and neutron pulse shape]{Comparison between the average normalized pulse shape
of IBD (denoted in blue) and neutron (denoted in red) events in the center of LENA with $\mathrm{E_{vis}=9.2\,MeV}$.}
\label{fig:pulse_shape_neutron}
\end{figure}

Figure \ref{fig:pulse_shape_neutron} shows a comparison between the average pulse shapes of neutron and IBD events in the center of the detector.
The average visible energy of the events was 9.2\,MeV, which corresponds to a $\mathrm{\bar\nu_e}$ with 10\,MeV. Compared to the small
scale laboratory experiment, the difference between the two pulse shapes is much smaller. The reason for this effect is that a large fraction
of the emitted photons are scattered before they are detected by a PMT, which distorts the pulse shape. Nevertheless, a clear difference between
the two pulse shapes is still visible, that can be used to identify fast neutron and atmospheric NC events.

For the pulse shape analysis, two different methods were used. In the tail-to-total method, the photon signal is integrated over two different intervals.
One interval includes the complete pulse, the so called total interval, and the other one encompasses the last part 
of the signal, the tail interval. Subsequently, the ratio between the tail and the total interval is calculated.

Another more complex method to discriminate between two different particle types 
is the gatti method \cite{gatti}.
First of all, the average pulse shapes for the two particles are calculated from events where
the particle type is known. Subsequently, a set of weights $\mathrm{P_i}$ is calculated from the two pulse shapes:
\begin{equation}
\mathrm{P_i=\frac{\alpha_i-\beta_i}{\alpha_i+\beta_i}} \ ,
\label{eq:gatti_weight}
\end{equation}
where $\alpha_i$/$\beta_i$ are the normalized number of photons detected in bin i of the 
$\alpha$/$\beta$ signal.
In the last step the so called gatti parameter $\mathrm{G_S}$ for a normalized signal $\mathrm{S_i}$
from an unknown particle is calculated:
\begin{equation}
\mathrm{G_S=\sum_i P_i S_i} \ .
\label{eq:gatti_parameter}
\end{equation}
Due to the weigths $\mathrm{P_i}$, which are determined by the average pulse shapes
of the two particles, the value of the gatti parameter depends on the particle type.

\begin{figure}[htbp]
\begin{center}
	\begin{minipage}[t]{0.48\textwidth}
		\includegraphics[width=1\textwidth]{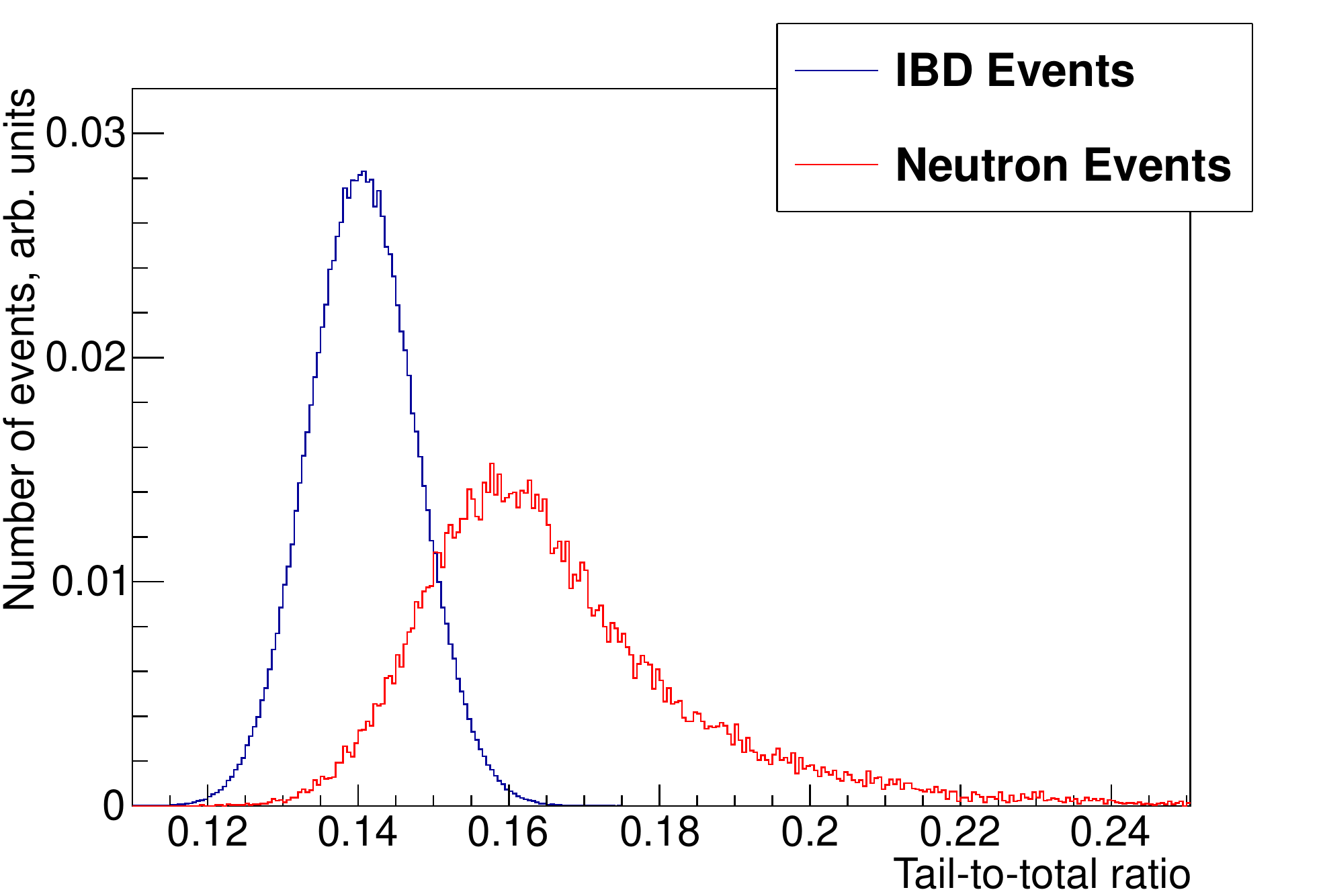}
		
	\end{minipage}
	\hskip 0.5cm
	\begin{minipage}[t]{0.48\textwidth}
		\includegraphics[width=1\textwidth]{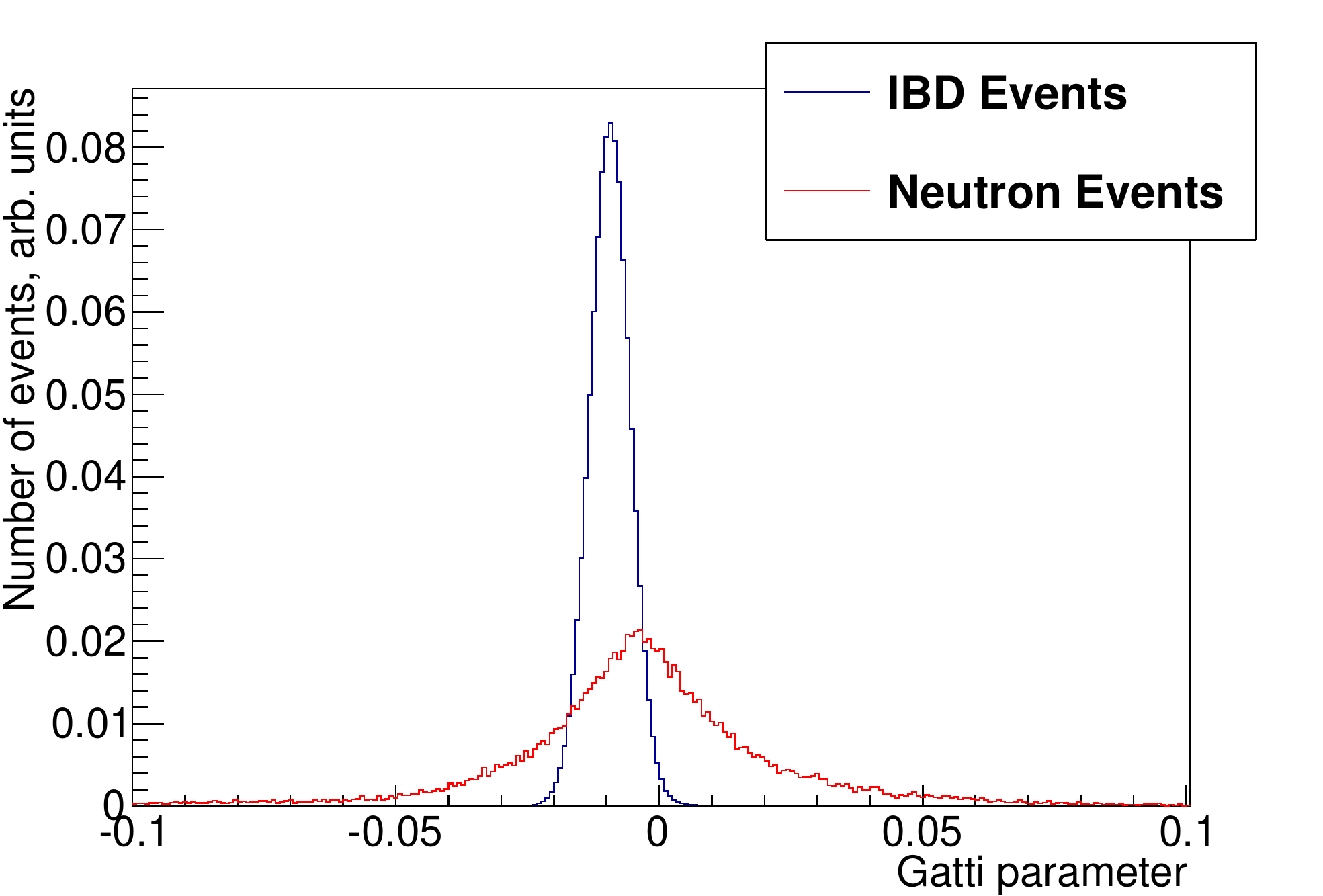}
	\end{minipage}	
\end{center}	
\caption[TTR and Gatti distribution]{The tail-to-total ratio (upper plot) and gatti parameter (lower plot) distribution for IBD (denoted in blue) and neutron events (denoted in red) in the center of LENA
with $\mathrm{E_{vis}=9.2\,MeV}$.}
\label{fig:ttr_gatti}
\end{figure}

Figure \ref{fig:ttr_gatti} shows the tail-to-total ratio and gatti parameter distribution for IBD and neutron events. For the tail-to-total ratio distribution,
a clear difference between IBD and neutron events is visible, while there is also an overlap. Hence, an efficient discrimination is only possible
for a low acceptance of IBD events. The gatti parameter distribution shows a large overlap between IBD and neutron events. The reason for this is that
there are large fluctuations of the indivual neutron pulse shapes, as neutrons can make elastic scattering reactions on protons as well as inelastic
scattering reactions on carbon, which lead to a different pulse shape. Hence, the gatti weights are not optimized for each of the possible reactions. Nevertheless,
it was found that the gatti method performs well on some pulses, where the tail-to-total method does not perform well. Hence, the overall discrimination
efficiency can be enhanced by applying both the tail-to-total and the gatti method.

\begin{table}
\begin{center}
\begin{tabular}{|c|c|c|c|}
\hline
IBD       &  Atmospheric & Fast neutron & DSNB Signal\\
acceptance     & NC rate [10\,y] & rate [10\,y] & $\mathrm{\langle E_\nu \rangle=12\,MeV}$ \\                       
\hline
90.0\,\% & $\mathrm{378 \pm 2 (stat.)}$    & $\mathrm{8.6 \pm 0.5 (stat.)}$  &  40.2\\
80.0\,\% & $\mathrm{155 \pm 1 (stat.)}$    & $\mathrm{4.5 \pm 0.4 (stat.)}$  &  35.8\\
50.0\,\% & $\mathrm{34.4 \pm 0.5 (stat.)}$ & $\mathrm{2.1 \pm 0.3 (stat.)}$  &  22.4\\
40.0\,\% & $\mathrm{21.8 \pm 0.4 (stat.)}$ & $\mathrm{1.8 \pm 0.2 (stat.)}$  &  17.9\\
\hline
\end{tabular}
\end{center}
\caption[Atmospheric NC background rate after pulse shape cuts]
{The atmospheric NC and fast neutron background depending on the IBD acceptance of the used pulse shape cut for 44\,kt fiducial volume.
For comparison, the corresponding DSNB event rates for $\mathrm{\langle E_\nu \rangle=12\,MeV}$ are also denoted.}
\label{tab:nc_bg_psd}
\end{table}

Table \ref{tab:nc_bg_psd} shows the atmospheric NC and fast neutron background rate, depending on the acceptance for IBD events of the used pulse shape cut
for 44\,kt fiducial volume. In case that a strict pulse shape cut is used, which only accepts 40\,\% of all IBD events, the atmospheric NC background is reduced by two orders of magnitude to $\mathrm{21.8 \pm 0.4 (stat.)}$ events per 10\,y. This cut also reduces the fast neutron rate to $\mathrm{1.8 \pm 0.2 (stat.)}$ events per 10\,y, so that no further fiducial volume cut needs to be applied.

\section{Detection Potential}
\label{sec:dsnb_detection}

\begin{table}
\begin{center}
\begin{tabular}{|c|c|}
\hline
Background source    &  Rate [10\,y]\\
\hline
Reactor neutrinos & 2.0\\
Atmospheric $\bar\nu_e$ & 2.2\\
$^9$Li $\beta^- -n$& $<0.01$\\
Fast neutrons & 1.8\\
Atmospheric NC&21.8\\
\hline
Sum & 27.8\\
DSNB ($\mathrm{\langle E_\nu \rangle=12\,MeV}$) & 17.9\\
\hline
\end{tabular}
\end{center}
\caption[Background rates in the DSNB detection window]
{The expected background rates in the DSNB detection window after applying the pulse shape and the fiducial volume cut.}
\label{tab:background_rates}
\end{table}

Table \ref{tab:background_rates} summarizes the contribution of the different background sources after applying the pulse shape and the fiducial volume cut.
Overall, 27.8\, background events per 10\,y are expected inside the detection window, while the predicted
DSNB rate ranges from 17.9 ($\mathrm{\langle E_\nu \rangle=12\,MeV}$) to 35.2 ($\mathrm{\langle E_\nu \rangle=21\,MeV}$) events per 10\,y.

A prerequisite for a positive detection of the DSNB is that the expected background rate is determined with a high precision. The reactor and
atmospheric $\mathrm{\bar\nu_e}$ rate can be extrapolated from the measured rate outside the detection window. As the fast neutron rate decreases with the radius of the fiducial volume, it can be determined by analyzing the dependence 
of the event rate on the radius of the reconstructed position.

Measureing the atmospheric NC event rate is challenging, as the efficiency of the applied pulse shape cut
must be known with high precision. A possible option is to look for events in the center of the detector, where two neutron captures
were detected, as these events can only be due to atmospheric neutrino NC reactions. While the efficiency for IBD-like atmospheric NC events
cannot be calculated directly from these events, they can be used to validate the Monte-Carlo simulation. Subsequently, the 
atmospheric NC event rate can be calculated with this validated Monte-Carlo simulation.

It is not clear how precise the background measurement will be. Hence, two scenarios are dicussed in following. An optimistic scenario of 5\,\% background uncertainty
and pessimistic one of 25\,\% background uncertainty. In order to assess the expected significance for a positive detection of the DSNB, it was assumed
that the number of detected events equals the sum of the expected signal and background rate.
From the number of detected events, background rate and background uncertainty, the confidence interval for the DSNB rate is calculated
according to \cite{t_rolke}. By increasing the size of the confidence interval such that
the lower limit is almost zero, the significance of the DSNB detection is calculated. 
Table \ref{tab:dsnb_det_pot_rolke} shows the resulting detection significance.
In the optimistic case that the background uncertainty is 5\,\%, the DSNB can be detected
with 3\,$\sigma$ significance in all cases. If the background is known with 25\,\% precision, a 3\,$\sigma$ detection is only possible for 
$\mathrm{\langle E_\nu \rangle>=18\,MeV}$. Hence, the background needs to be well understood in order to detect the DSNB.

\begin{figure}[!htbp]\centering\includegraphics[width=0.49\textwidth]{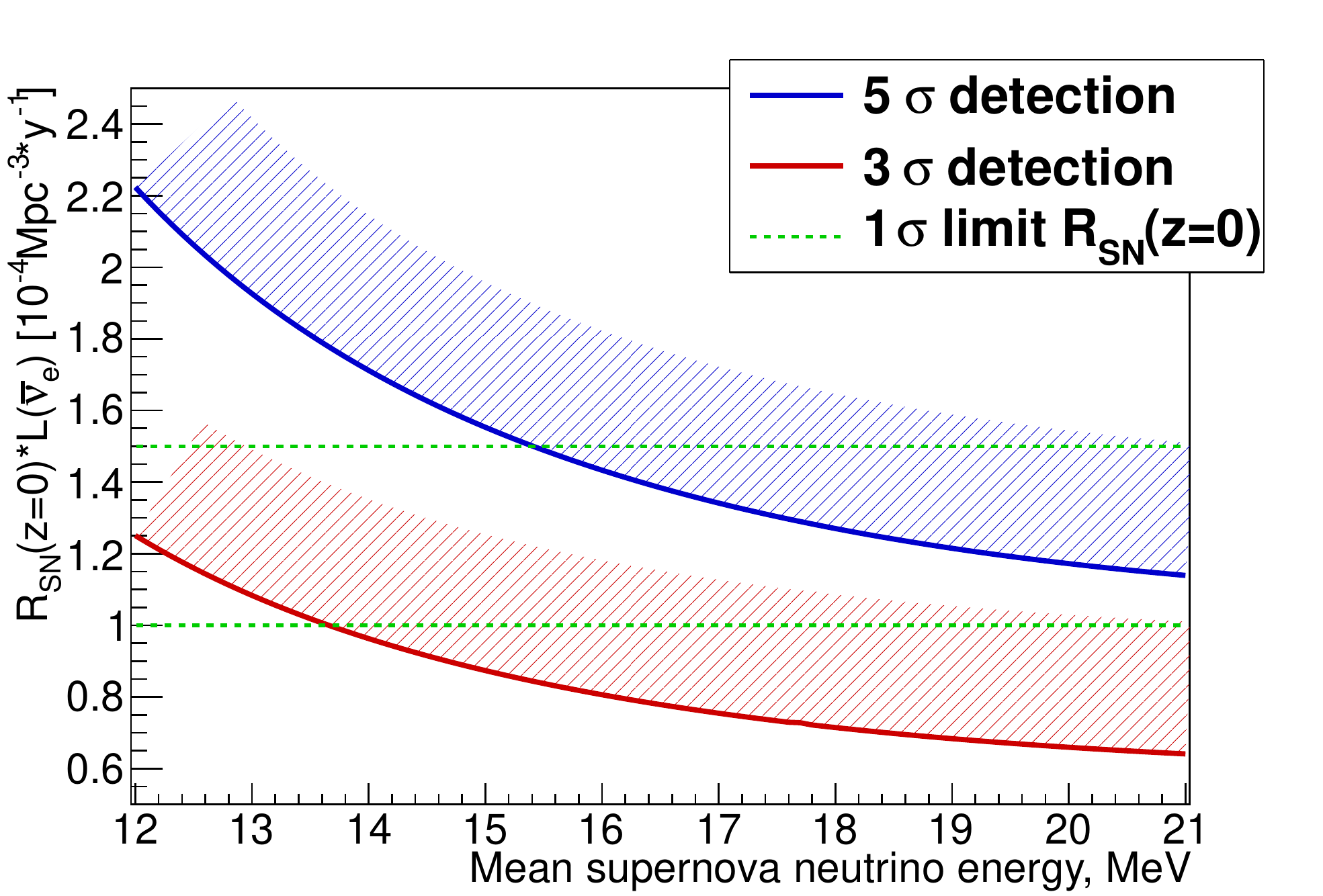}
\caption[DSNB detection potential]
{The 3\,$\sigma$ and 5\,$\sigma$ detection potential for the DSNB, depending on the supernova rate $\mathrm{R_{SN}(z=0)}$ and the mean supernova neutrino energy for 5\,\% background uncertainty and 10 years measuring time.
For comparison, the current $1\,\sigma$ confidence interval for $\mathrm{R_{SN}(z=0)}$ (green dashed lines) \cite{dsnb_beacom}
is depicted.}
\label{fig:detection-potential}
\end{figure}

Figure \ref{fig:detection-potential} shows the DSNB detection potential, depending on the supernova rate $\mathrm{R_{SN}(z=0)}$ and the mean supernova neutrino energy for 5\,\% background uncertainty. For $\mathrm{\langle E_\nu \rangle \ge 14\,MeV}$, a 3\,$\sigma$ detection is possible for the whole currently
predicted range for the supernova rate. If the true supernova rate is in the upper part of the predicted range, even a 5\,$\sigma$ detection is possible
for $\mathrm{\langle E_\nu \rangle \ge 16\,MeV}$.

\begin{table}
\begin{center}
\begin{tabular}{|c|c|c|}
\hline
$\mathrm{\langle E_\nu \rangle}$ &5\,\% background & 25\,\% background \\
& uncertainty& uncertainty \\
\hline
12\,MeV & $3.0\,\sigma$ & $1.9\,\sigma$\\
15\,MeV & $4.0\,\sigma$ & $2.6\,\sigma$\\
18\,MeV & $4.9\,\sigma$ & $3.1\,\sigma$\\
21\,MeV & $5.4\,\sigma$ & $3.5\,\sigma$\\
\hline
\end{tabular}
\end{center}
\caption[DSNB detection significance including a possible background uncertainty]{The expected detection significance after 10 years for different DSNB models
with $\mathrm{\langle E_\nu \rangle}$ ranging from 12\,MeV to 21\,MeV, assuming that the expected background rates are known
with 5 and 25\,\% precision, respectively.}
\label{tab:dsnb_det_pot_rolke}
\end{table}

In case that the number of detected events does not exceed the background rate, the current limit on the DSNB flux of the Super-Kamiokande experiment could 
be significantly improved. Assuming that 28 events are detected after 10\,y ($\mathrm{N_ {det}=\langle N_{bg} \rangle}$) and that the background is known
with 5\,\% uncertainty, the upper limit on the DSNB flux above 17.3\,MeV would be $\mathrm{0.4\,\bar\nu_e\, cm^{-2}s^{-1}}$ for $\mathrm{\langle E_\nu \rangle=18\,MeV}$, which is a factor of about 8 below the current limit \cite{superK_dsnb}.

\begin{figure}[!htbp]\centering\includegraphics[width=0.49\textwidth]{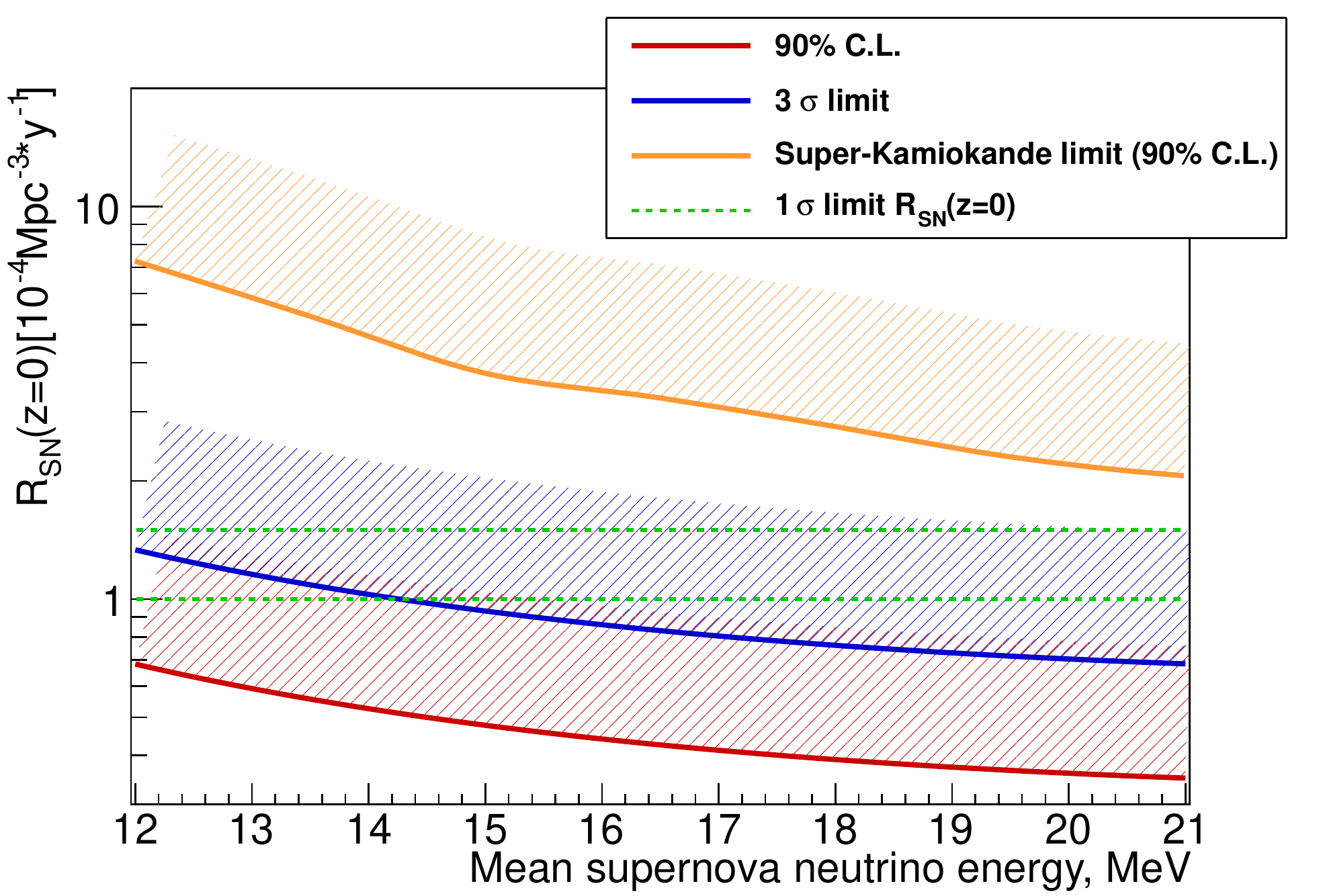}
\caption[Exclusion contours for the supernova rate $\mathrm{R_{SN}(z=0)}$ and the mean supernova neutrino energy if no DSNB signal is detected]
{The $3\,\sigma$ (depicted in blue) and 90\,\% C.L. (depicted in red) exclusion contours
for the supernova rate $\mathrm{R_{SN}(z=0)}$ and the mean supernova neutrino energy,
assuming 5\,\% background uncertainty and no detected DSNB signal ($\mathrm{N_ {det}=\langle N_{bg} \rangle}$).
For comparison, the current $1\,\sigma$ confidence interval for $\mathrm{R_{SN}(z=0)}$ (green dashed line) \cite{dsnb_beacom}
and the 90\,\% C.L. exclusion contours of the Super-Kamiokande experiment are depicted.}
\label{fig:exclusion_limit}
\end{figure}

Figure \ref{fig:exclusion_limit} shows the exclusion contours for the supernova rate $\mathrm{R_{SN}(z=0)}$ and the mean supernova neutrino energy,
assuming 5\,\% background uncertainty and that no DSNB signal was detected. 
Independent of the mean supernova neutrino energy, the whole currently predicted range for the supernova rate would be covered by the 90\,\% exclusion
limit. The exclusion would be even at the 3\,$\sigma$ level if $\mathrm{\langle E_\nu \rangle \ge 15\,MeV}$.
Hence, if no excess above the expected number of background events
is found in LENA, all current standard DSNB models would be ruled out with more than 90\,\% C.L.

\section{Conclusions}
\label{sec:con}

Due to its large target mass, the proposed LENA detector will be sensitive to
the still undetected Diffuse Supernova Neutrino Background (DSNB).
Indistinguishable background from reactor and atmospheric electron antineutrinos limit the detection window
to neutrino energies from 9.5\,MeV to 25\,MeV. Depending on the mean supernova neutrino energy, about 50 to 100 events per 10\,y are expected in this 
energy region. The most crucial background arises from neutral current reactions of atmospheric neutrinos, which can mimic the IBD event signature.
It surpasses the DSNB signal by more than one order of magnitude. This background can
be significantly reduced by a pulse shape analysis to $\mathrm{21.8\pm0.4(stat.)}$ events per 10\,y.
Although this cut also reduces the DSNB event rate by 60\,\%, a signal to background ratio of around 1 is expected.

Assuming that the background rate is known at 5\,\% uncertainty, LENA can detect the DSNB with $3\,\sigma$ significance after 10\,y. In case
that no signal is found, the current upper limit on the DSNB flux can be improved by a factor of 8. Furthermore, all current standard DSNB  models
could be ruled out with more than 90\,\% C.L.

\section*{Acknowlegdements}

This research was supported by the DFG cluster of excellence 'Origin and Structure of the Universe' (Munich) and 'PRISMA' (Mainz).

\bibliography{lit}
\end{document}